\documentclass{ws-procs9x6}
\usepackage{color}

\begin{document}

\title{Perspectives of Nuclear Physics 
}

\author{Amand Faessler
}

\address{University of Tuebingen \\
Institute for Theroetical Physics\\ 
D-72076 Tuebingen, Germany}


\maketitle

\abstracts{
The organizers of this meeting have asked me to present perspectives of nuclear
physics. This means to identify the areas where nuclear physics will
be expanding in the next future. In six chapters a short
overview of these areas will be given, where I expect that nuclear physics will develop quite
fast: }

\begin{enumerate}
\item Quantum Chromodynamics and effective field theories in the confinement
region.
\item Nuclear structure at the limits.
\item High energy heavy ion collisions.
\item Nuclear astrophysics.
\item Neutrino physics.
\item Test of physics beyond the standard model by rare processes.
\end{enumerate}

\abstracts{ After a survey over these six points I will pick out a few topics
where I will go more in details. There is no time to give for all six points
detailed examples. I shall discuss the following examples of the six topics
mentionned above:}

\begin{enumerate}
 \item The perturbative chiral quark model and the nucleon $\Sigma$-term. 
 \item VAMPIR (Variation After Mean field Projection In Realistic model spaces
 and with realistic forces) as an example of the nuclear structure renaissance.
 \item Measurement of important astrophysical nuclear reactions in the Gamow
 peak.
 \item The solar neutrino problem.
\end{enumerate}

\abstracts{ As examples for testing new physics beyond the standard model by
rare processes I had prepared to speak about the measurement of the electric
neutron dipole moment and of the neutrinoless double beta decay. But the time
is limited and so I have to skip these points, although they are extremely
interesting.}

\section{The view from the top}
\subsection{Quantum Chromodynamics and effective field theories in the confinement
region}

We all are convinced that Quantum Chromodynamics (QCD) is the correct theory of
the strong interaction. Thus all considerations about nuclear physics should be
starting from QCD. But no-one will request that one is calculating the structure
of $^{208}Pb$ using QCD. But one should calculate with lattice QCD the
properties of the baryon resonances and the mesons. This is done presently with
increasing success. The main problem in lattice QCD is to implement
chiral symmetry, which is an important property of QCD. My friends working about
QCD on the lattice tell me, that this problem has been solved within the last
two years, at least in principle. But it is still numerically very difficult to
reduce the masses of the quarks and the pion to reach the chiral limit. But one
can for example take QCD lattice data and extrapolate it them with the help of 
chiral perturbation theory to the chiral
limit. 

Chiral Perturbation Theory (ChPT) can be proved \cite{leut} to be equivalent to
QCD in a low energy limit. On the other side Weinberg and others
\cite{wein,weinb,hadji} did show that specific constituent quark Lagrangians are
equivalent to an effective description of hadrons and their interaction. So one
can expect that there exist effective field theoretical Lagrangians of quark models,
which are equivalent in the low energy limit to QCD \cite{wein,weinb,hadji,sal}.
But such an equivalence has not yet been proved.

The focus is presently on the structure of the nucleons (electromagnetic form
factors) and on the strangeness in the nucleon (spin structure functions). 

In the last years one has also learned to partition a process in a hard and soft
part. For the hard part one can use perturbative QCD and for the soft part one can
use models or experimental form factors. Such a process is shown with its lowest
order diagramm in figure 1. 

\begin{figure}[h]
\centerline{\epsfxsize=3.9in\epsfbox{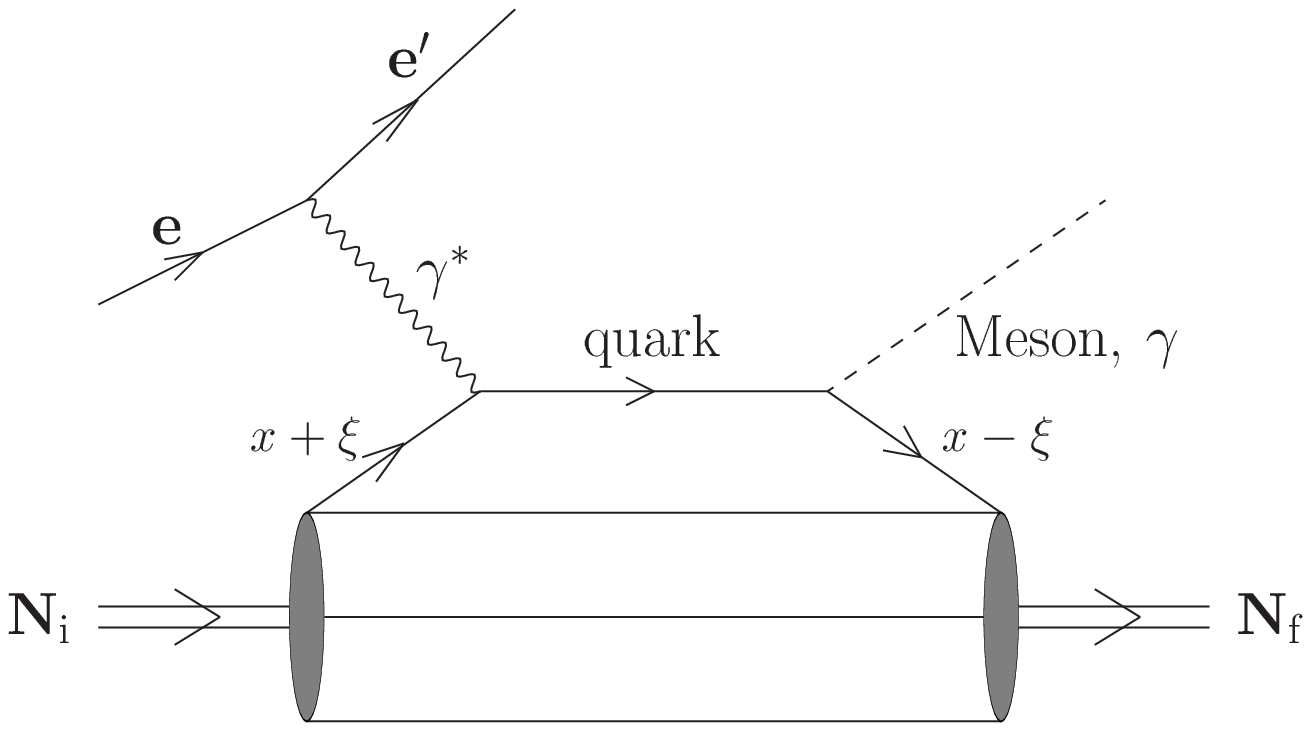}}   
\caption{``Handbag'' diagram for the electro production of a meson or a
$\gamma$ quant. This lowest order diagramm partitions electro production of a
meson or a gamma quant into a hard piece, involving one parton  with the initial
fractional momentum $x + \xi$ and the final fractional momentum $x - \xi$. The
hard process of the electron scattering with the parton and the meson or the
gamma production can be treated by QCD perturbation theory. The soft part of
this reaction is parametrized by an off-forward form factor which not only
depends on the momentum transfer $Q^2$ and on the momentum fraction of the
interacting parton $x$, but on the momentum fraction of the parton in the
initial state of the nucleon $x + \xi$ and on the momentum fraction of the parton in the
final state of the nucleon $x + \xi$. \label{inter}}
\end{figure}
 
One gets additional information by polarising the proton and the electron or the
muon. Recently at HERMES and at COMPASS one discusses the situation were the
proton is transversally polarized (transversity). Since the gluons like the
photons have their spin in or against to the flight direction, they cannot
contribute to the transversally polarized proton spin. Since the sea quarks and
antiquarks are produced by the gluons, they are also not contributing to the
proton spin, if the proton is polarized perpenticular to the beam axis. Thus
the spin structure function for a proton polarized in this way must be simpler
than for a longitudinal polarized proton. 

An open problem are hybrids and glueballs: hybrids are hadrons composed
partially by quarks and partially by gluons. Exotic mesons would be composed of
a quark and an antiquark and one or several gluons. They can have quantum numbers
which are not allowed if they are built soly by a quark and an antiquark. 

Glueballs are particles built only by gluons. The LEAR collaboration found a
fith neutral scalar ``meson'' with the quantum numbers $0^+$. For
neutral scalar mesons only $a_0 (1450)$, $K^{o*} (1430)$, $f_0 (1370)$ and the $f_0
(1710)$ should exist. Hence
the $f_0 (1500  MeV) $ could be a glueball. Indeed this state is only
weakly decaying into two photons. A pure glueball cannot decay into two photons,
since it does not contain any charged particles. But the weak decay into two
photons already indicates that the $f_0 (1500)$ can at the best be a mixture of
a glueball with a quark - antiquark configuration. This is also suggested by
theoretical calculations \cite{gu,am}.

In spite of the strong hints for the existence of exotic mesons and of glueballs
they have not been detected out of any doubt. 


\subsection{Nuclear Structure}

As mentioned already above, no-one will request that nuclei like $^{208}Pb$ are
calculated starting from QCD. But we would like to understand the
nucleon-nucleon interaction within the framework of QCD. Presently we have only
QCD inspired models describing the nucleon-nucleon interaction
\cite{NN-Quark,kukulin}. 

From the bare nucleon-nucleon interaction one has to derive in a second step the
effective nucleon-nucleon force for a given model space, in which the nuclear
many-body problem is solved. This deduction of the effective nucleon-nucleon
interaction from the bare nucleon-nucleon force has been studied since the
60ies. But the problem has not yet been solved quantitatively. A different
question is then the derivation of the simple very successful nuclear models
from the nucleon degrees of freedom and the effective interaction. 

Presently we have a renaissance of experimental nuclear structure studies and
the corresponding theoretical investigations. This renaissance is on one side
due to the possibility of radioactive beams: accelerators like SPIRAL in Caen,
RIKEN in Japan, the GSI in Darmstadt (existing and planned facilities), the
Rare Ion Accelerator (RIA) discussed in the States and the smaller activities in
Catania and Legnaro in Italy. On the other side this renaissance is due to new
detector arrays like Euroball and the Gamma Sphere and plans of even more
advanced gamma-ray detectors. 

One aim is to study the shell structure of nuclei at the limits of stability for
larger proton numbers Z and neutron numbers N. In the last years one found the
celebrated neutron halos starting in $^{11}Li$ and proton skins, which are
smaller than the neutron halos due to the high Coulomb barrier. One detected
also new types of collective states: D\"onau, Frauendorf and others
\cite{doenau,huebel} predicted
magnetic rotations in spherical nuclei. H\"ubel and Clark \cite{huebel} found these states
experimentally. The essence of these magnetic rotations are spherical nuclei,
mostly with some proton holes just below a magic number and some neutron
particles just above. Since particles and holes are repelling each other, the
orbits of neutrons and protons try to have an overlap as small as possible (see
figure 2). 

\begin{figure}[h]
\centerline{\epsfxsize=3.0in\epsfbox{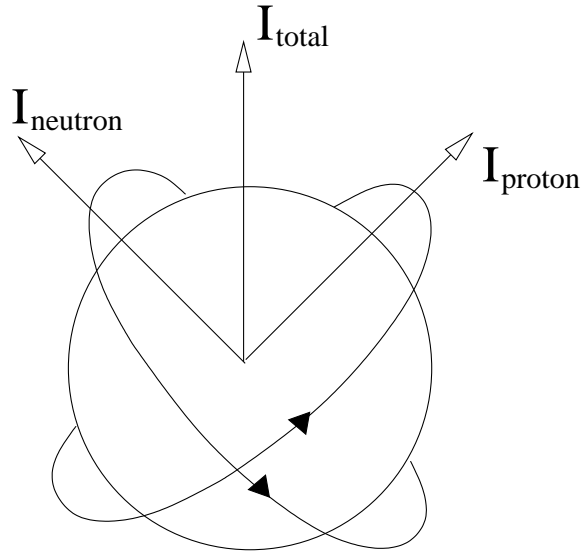}}   
\caption{Magnetic rotational nucleus. The protons are normally hole states
below a closed shell while the neutrons are particle states above a closed
shell. Particles and holes are repelling each other, so that the orbits of proton
holes and the neutron particles try to arrange with the least possible overlap.
With increasing angular momentum the shear of the proton and the neutron angular
momenta is closing and the energy band looks rotational. Since the shape of the
nucleus is almost spherical, one has only very weak electric quadrupole
transitions, but strong magnetic dipole M1 transitions. When the proton and the
neutron angular momenta start to align, the M1 transitions are getting weaker,
since the magnetic moment of the protons is along the proton angular momentum
and the magnetic momentum of the neutrons is opposite to the angular momentum of
the neutrons.  \label{inter}}
\end{figure}

Figure 3 shows such magnetic rotational bands with strong M1 transitions
measured in the $^{54}Fe (^{56}Ni, \alpha 4 p)$ at a Ni kinetic energy of
$E_{kin} = 243 MeV$ by Jenkins et al. at Berkely \cite{jen} in $^{102}Cd$ and
$^{104}Cd$. 

\begin{figure}[h]
{\epsfxsize=3.0in\epsfbox{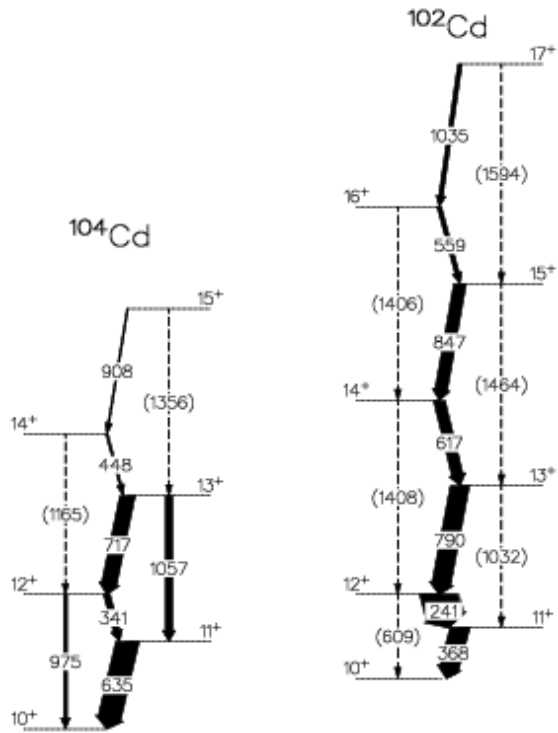}}   
\caption{Magnetic rotational spectra of $^{104}Cd$ and $^{102}Cd$.
The protons form two holes in $g_{9/2}$ below the magic number 50 and the 
neutrons six particles in the $d_{5/2}$ and the $g_{7/2}$ shells. 
The width of the arrows between states with neighbouring angular momenta 
indicate the strength of the magnetic dipole M1 transitions, while 
$\Delta$I = 2 transitions are of E2 nature. One sees that with increasing 
angular momenta the magnetic dipole M1 transitions are reduced and that 
the electric quadrupole E2 transitions are always weaker than the M1 
intensities. (Figure taken from D. G. Jenkins et al., arXiv: nucl-ex/0007004 vl.)}
\end{figure}

Superdeformed nuclei with an axis ratio 2:1 have been already found in the
80ies. Hyperdeformed nuclei with an axis ratio 3:1 have been predicted by
Strutinsky type calculations since a long time. But in spite of several
announcements that such hyperdeformed nuclei have been found, such an axis ratio
has been not yet clearly established. 

The shell model technology made large progress over the last years
\cite{strass}. One is now able to diagonalized matrices of the many-body
hamiltonian of around 100 millions times 100 millions. This allows to treat a many-body
Hamiltonian exactly in the pf-shell. The Green's function Monte Carlo approach
\cite{gran} allows to solve the nuclear many-body problem exactly without any
restriction for up to 8 or even 10 nucleons. A different approach is used in
Tuebingen \cite{vamp} and Tokyo \cite{ots}. In the VAMPIR (Variation After Mean
field Projection In Realistic model spaces and with realistic forces) and in the
Monte Carlo shell model \cite{ots} one is selecting the many-body basis states so
carefully, that a few states are producing similar good results as a shell model
diagonalization of several million configurations. 

Relativistic nuclear structure approaches \cite{ring} have the advantage, that
they are able to describe the spin-orbit part of the self-consistent
nucleon-nucleus potential quantitatively \cite{ring}.


\subsection{High energy heavy ion collisions}

One of the main aims of studying ultrarelativistic heavy ion collisions at CERN
with the SPS, at Brookhaven with RHIC and in the future also with the ALICE
detector at the LHC (Large Hadron Collider)/CERN is to study (and find) the
phase transition from nuclear to quark matter. 

The detection of the quark-gluon plasma has already been announced on February
10th, 2000 by CERN. For such a phase transition we have strong circumstantial
evidence, but not a definite proof. The strong evidence is
three fold: 

\begin{itemize}
\item[(i)] The suppression of the $J/\psi$ production in ultrarelativistic heavy
ion collisions at SPS/CERN relative to the proton-proton reaction. 
\item[(ii)] An increase of the number of hadrons with strangeness content in ultrarelativistic
heavy ion collisions relative to the proton-proton reaction at the same energy
per nucleon.
\item[(iii)] Hadrons are produced in ultrarelativistic heavy ion collisions in
chemical equilibrium. This means that their relative abundance can be described
by two parameters: a temperature and a chemical potential. Simulations which
allow only the hadronic side, cannot reproduce well enough the hadron production in heavy
ion collisions in chemical equilibrium. Thus one believes that the hadrons are
produced during the condensation from the quark-gluon plasma into the hadronic
phase in equilibrium characterized by the condensation temperature and the
chemical potential for the baryon density. 
\end{itemize}

At the ultrarelativistic heavy ion
collisions at LHC (Large Hadron Collider at CERN) one should be able to
study the melting of the vacuum. This means the restoration of chiral symmetry
at high temperatures where one expects that the quark condensate $\langle \bar{q}
q \rangle$, which has at zero temperature a finite expectation value, disappears
$\langle \bar{q} q \rangle = 0$. The GSI plans to build a heavy ion
collider with about 25 GeV per nucleon to study a possible phase transition from
hadronic to quark matter at high baryonic density.

\begin{figure}[ht]
\centerline{\epsfxsize=3.9in\epsfbox{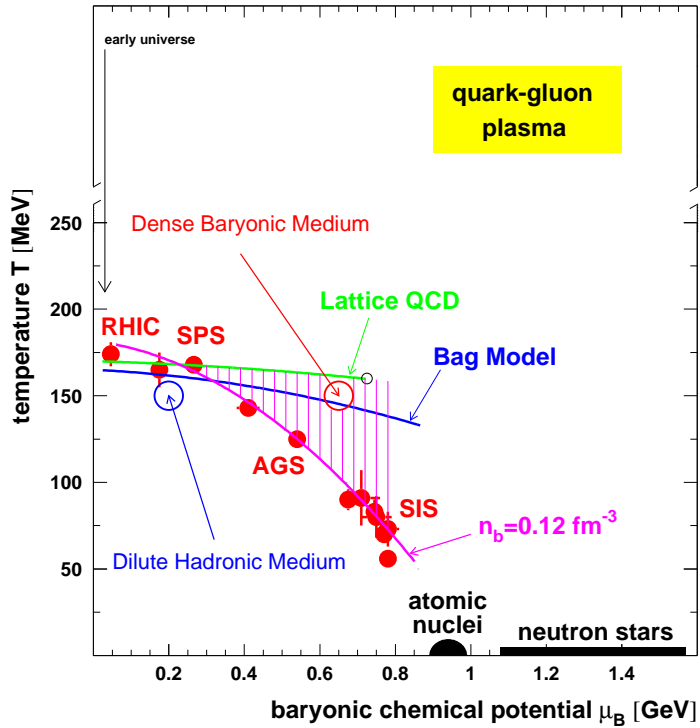}}   
\caption{ Phase diagram of the temperature against the baryon chemical
potential. One line indicated ``lattice QCD'' shows the phase transition obtained in
lattice QCD from nuclear to quark matter. The other lines with the dots and the
bars show the temperatures and the chemical potentials determined in different
heavy ion reactions and at different accelerators (RHIC, SPS, AGS and SIS)
extracted from hadron multiplicities after heavy ion collisions. The baryon
chemical potential $\mu_B$ in atomic nuclei is roughly given by the nucleon mass
of 938 MeV. (Figure taken from P. Braun-Munzinger, J. Stachel, J. Phys. G 28,
(2002) 1971 - 1976.)\label{inter}}
\end{figure}

Figure 4 shows the temperature against the baryon chemical
potential $\mu_B$. The phase transition line expected from lattice QCD and the
temperatures and chemical potentials determined from the multiplicity of hadrons
produced in heavy ion collisions at different accelerators are displayed. 

At RHIC one found a quite surprising result: Quarks are moving through a nucleus
without loosing too much energy (color transparency), while in a quark-gluon
phase, due to open color a quark is frequently interacting and producing gluons.

Figure 5 shows the number of neutral pions produced in heavy ion collisions
(mainly Pb on Pb at CERN and Au on Au at RHIC) relative to the number of neutral
pions in proton-proton collisions scaled to the collisions of the indicated two
heavy nuclei. A surprising result is, that this ratio behaves quite differently
for the collisions at the SIS and at RHIC. The value 1 would be obtained, if
nothing new happens in heavy ion collisions compared to proton-proton reactions.
That the number of neutral pions in Au on Au collisions is below unity at RHIC
might indicate that the quark-gluon plasma is reached at RHIC. But it would also
suggest that at CERN one still might have by large the color transparency
situation with high energy jet. But the interpretation of these data is not
obvious. But it is clear that something
dramatically happens between the SIS and the RHIC energies. 

\begin{figure}[h]
\centerline{\epsfxsize=3.9in\epsfbox{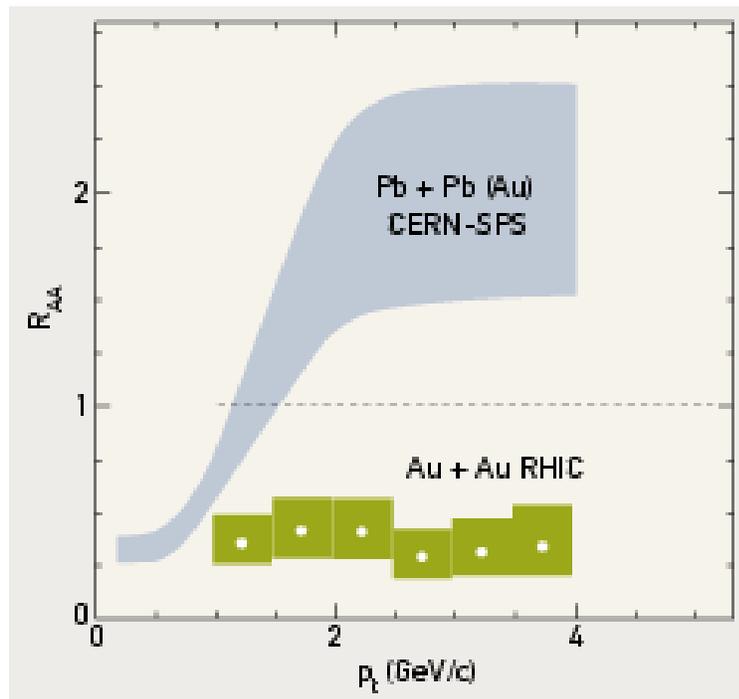}}   
\caption{Number of neutral pions in heavy ion collisions compared
to the number of neutral pions in the proton-proton reaction scaled to the heavy
ion reaction as a function of the perpendicular momentum of the emitted
particles. Without any new effects in heavy ion collisions this ratio should be
unity. That this ratio behaves differently for the data from the SIS/CERN and
for the data from RHIC/Brookhaven indicates that something dramatically happens
between the SPS cm energy of 17 GeV per nucleon and for RHIC cm energy of up to
200 GeV per nucleon in the center of mass system. \label{inter}}
\end{figure}

An open question is also the nature of the phase transition from nuclear to
quark matter: is it a first or second order phase transition or is there a critical 
point
where the phase transition on the right-hand side is of
first oder and on the left-hand side is smooth and of second order. A
first order phase transition would indicate large fluctuations in the growth of
the hadronic droplets in the big bang. This would immediately introduce large
inhomogenuities of hadronic matter in our universe. Such large fluctuations have
been seen by Boomerang and by Maxima in the cosmic background radiation. 


\subsection{Nuclear astrophysics}

The measurements of Boomerang and Maxima of the cosmic background radiation show
in their multipole decomposition of the fluctuations that the total energy
density in our universe should be very close to the critical density, which
produces a flat space. The hights of the second maximum indicates that hadronic
matter is of the order of 5 $\%$ of the saturation density. Since dark
matter is about 30 $\%$, it leaves 65 $\%$ of the saturation density for
vacuum enery or for a cosmological constant. This is in agreement with the
relative abundance of the light elements produced in the big bang, which
indicates also hadronic matter of the order of 5 $\%$. The motion of the
galaxies and the motion of stars in the halo of our galaxy suggest dark matter
of about 30 $\%$ of the saturation density. Visible matter in the
universe is below or of the order of 1 $\%$. A large vacuum energy or a cosmological
constant is also indicated by the observation of supernovae explosions at very far
distances. 

What is the nature of dark matter and of the vacuum energy or the cosmological
constant? Is dark matter composed of weakly interactive massive particles
(WIMP's) as suggested by the DAMA experiment in the Gran Sasso \cite{berna}. Or can
dark matter be explained trivially by a modification of Newtons law at small
accelerations? \cite{dark}.

To study the formation of elements in stars, one needs to know a large number of
cross-sections at extremely small energies (see discussion below). 

Medium heavy and extremely heavy nuclei are produced in the s-, p- and in the
r-process. 

In the s-process one needs a location where one produces neutron fluxes of about
$10^{8}$ neutrons per $cm^2$ and per second. In this case the beta decay
($\beta^-$) probability is faster than the $(n, \gamma)$
reaction. Thus the s-process follows the stable nuclei. In the r-process one
needs neutron fluxes which are about 14 orders of magnitude larger. The time for
the fusion of a neutron to a nucleus in a $(n, \gamma)$ reaction is much smaller
than the time for the $\beta^-$ decay. One therefore forms nuclei with an
extremely large neutron number up to the next magic shell, where the neutrons
get unbound. There then they decay back to the stable nuclei in $\beta^-$
processes. 

The p-process happens in double stars with a white dwarf or a neutron star
and a red giant. Hydrogen is flowing from the red giant to the white dwarf into
an accretion disk and
is exploding after enough hydrogen has been acclomerated on the  surface of the
white dwarf to ignite the fusion of hydrogen with the nuclei at the surface of
the white dwarf. This produces very proton rich nuclei, which decay by $\beta^+$
reactions back to the stable nuclei. 

An open problem are supernovae explosions. They are triggered after large stars
of about 10 to 20 times  solar masses have formed an inner Fe and Ni core. The
collaps of the stars starts after one reaches the condition

\noindent
\begin{equation}
m_e c^2 + \varepsilon_F \ge (m_n - m_p) c^2 = 1.293 MeV\,.
\label{eq:1}
\end{equation}

If the rest mass of the electron and its Fermi energy $\epsilon_F$ is getting
larger than the mass difference between the neutron and the proton of 1.293 MeV
the inverse beta decay sets in. 

\noindent
\begin{equation}
e^- + p \rightarrow n + \nu_e\,.
\label{eq:2}
\end{equation}

With the inverse beta decay the electrons are disappearing out of the star and
the electron neutrinos can escape and so the Fermi pressure of the electrons is
reduced and the star starts to collaps. This collaps goes on until the inner
part of the Fe and Ni core has a density of about $10^{12} g/cm^3$. At this density
the neutrinos of the reaction of the inverse beta decay (2) cannot any more
escape and they trigger a shockwave which leads to the explosion of the star. But
in all the simulations one does not have enough energy in the neutrinos to
explode the star. This is mainly due to the fact that the inner core of Fe and
Ni with the density larger than $10^{12}g/cm^3$ is surrounded by a large sphere
of Fe and Ni of lower density through which the shock has to travel. The energy which is
needed to break up the Fe and Ni layer stalls the shock. 

Recently Langanke suggested, that the inverse beta decay (2) leaves much 
more electrons in the center of a collapsing star, than expected up
to now. Since the high density area, in which the electron neutrinos are trapped
has a radius proportional to the remaining electron density squared, the shock
wave has to travel through a much smaller layer of Ni and Fe nuclei. This could
be perhaps one of the reasons why indeed supernovae are exploding in reality and
not in simulations. But another reason could be that all the calculations until
now are either one or two dimensional. A three dimensional calculation would
allow much smaller eddies and by that the situation of the exploding star would
be quite different. Naturally there could also be missing some until now unknown
piece of microphysics. 


\subsection{Neutrino physics}

Neutrino physics is presently a very fast developing field. The development of
neutrino physics was in recent years mainly fuelled by the solar neutrino
problem. The solar neutrino puzzle can be solved assuming that the neutrinos are
massive and that the mass eigenstates are different from the production or the
weak or flavour eigenstates. All solar neutrino detectors, the Cl detector in
the homestake mine (Nobel price 2002 to Ray Davies), the Ga detectors GALLEX and 
SAGE, the KAMIOKA, the
Super-KAMIOKA (Nobel price 2002 to Masatoshi Koshiba) and also the SNO detectors indicate that neutrinos are missing.
The same result was obtained from the atmospheric neutrinos, produced by cosmic
radiations in the atmosphere. Muon neutrino produced at the opposite side of the
earth are disappearing to a large amount until they reach the Super-KAMIOKA
detector in Japan. The
Sudburry neutrino observatory (SNO) in Canada could solve the problem by
measuring at the same time the neutral and the charge current reactions of the
neutrinos, using heavy water. The two measurements (where the charge current
reactions were borrowed from Super-Kamiokande) allow to determine the total
neutrino flux of electron, muon and tauon neutrinos. This total neutrino flux
corresponds to the expected electron neutrino current from solar models. 

The neutrino oscillations request a small extension of the standard model:
neutrinos must have masses, which are different from each
other. Since the oscillations depend only on the neutrino mixing matrix of the
unitary transformation from the neutrino mass eigenstates to the flavour
eigenstates and on the squares of the differences of the neutrino masses, one
cannot determine an absolute mass scale. But (in the so-called natural
hierarchy) the mass difference between the lowest
and the second lowest neutrino masses must be of the order of $10^{-3}$ to
$10^{-2}$ eV, while the mass difference between the second and the third
neutrino is of the order of $10^{-1} $ to 1 eV.


\subsection{Test of new physics beyond the standard model}

The standard model of the electroweak (Glashow, Salam and Weinberg, 1968) and
the strong interaction (Quantum  Chromodynamics = QCD) is extremely successful.
But it has also 16 free parameters (six quark masses, three
mixing angles and one phase of the Cabbibo, Kobayashi and Maskawa unitary
matrix, three coupling constants and the three masses of electrons, muons and
tauons. These 16 parameters have to be supplemented by three neutrino masses,
three mixing angles and one phase of the neutrino mixing matrix). 

The standard model seems to be an effective field theory, which is
exactly valid at energies far below the Grand Unification scale of the order of
$10^{15}$ GeV. By embedding the standard model in a Grand Unified model or
in supersymmetric or even superstring models reduces the number of
parameters. In Grand Unified theories the electroweak and the strong interaction is
described by one single force with one coupling constant. 

Possible new physics from Grand Unification, supersymmetry (but probably not 
from superstrings) might be tested at ultrahigh energies at
the LHC or perhaps also at TESLA/DESY. But several aspects of
theories beyond the standard model can also be tested at low energies by looking
to rare events. 

Massive neutrinos with a unitary mixing between the mass and the flavour
eigenstates allow for example the conversion of muons into electrons on nuclei.
Figure 6 shows a possible diagram which contributes to this muon to electron
conversion on nuclei, which is studied experimentally e. g. at the
Paul-Scherrer-Institute (PSI).

\begin{figure}[h]
\centerline{\epsfxsize=3.1in\epsfbox{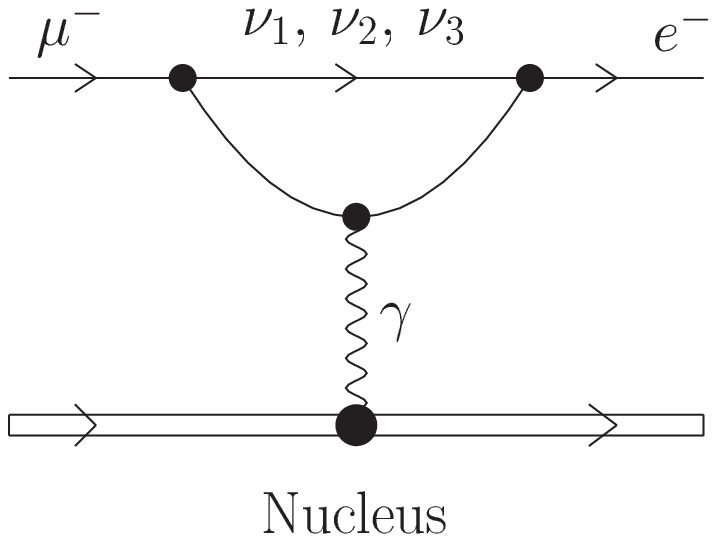}}   
\caption{ One diagram for the muon-electron conversion on nuclei
which is allowed by intermediate massive neutrinos and the unitary mixing of the
production eigenstates $\nu_e, \nu_{\mu}$ and $\nu_{\tau}$ and the neutrino mass
eigenstates $\nu_1, \nu_2$ and $\nu_3$. \label{inter}}
\end{figure}

Apart of the muon-electron conversion on nuclei, which tests only the hypothesis
of massive neutrinos and their mixing, one can test Grand Unification in the
neutrinoless double beta decay: 

\begin{figure}[h]
\centerline{\epsfxsize=3in\epsfbox{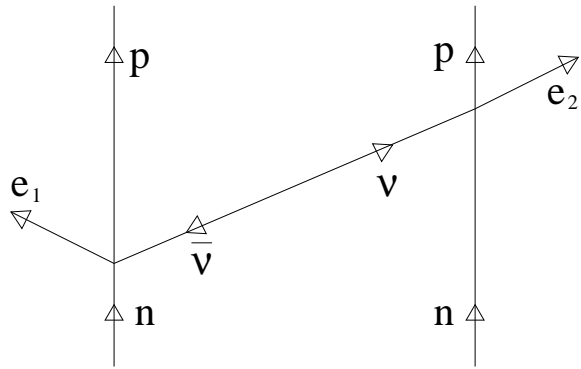}}   
\caption{This figure shows the simplest diagram of 
the double neutrinoless beta decay where
two neutrons in a nucleus are converted into two protons, which stay in the
nucleus. Two electrons are emitted and can be detected. Presently it seems that
the neutrinoless double beta decay has not been seen experimentally, although
the Heidelberg group of Klapdor and collaborators claims, that they have seen
the process.} 
\end{figure}

Figure 7 shows that the neutrinoless double beta decay is only possible if the
neutrino and the antineutrino are identical particles, which means the neutrino
must be a Majorana particle. This is predicted by most of the Grand Unified
theories. In addition for left-handed weak interaction theories, one has a
missmatch between the helicity of the emitted antineutrino and the 
absorbed neutrino. For massive neutrinos helicity is not a good quantum number
and thus the double beta decay is possible in Grand Unified theories with
massive Majorana neutrinos. Already now the lower limit of the lifetime for the
neutrinoless double beta decay of the order of $10^{25}$ years for $^{76}Ge$
\cite{klap} allows to restrict severely parameters of Grand Unified theories and
of the supersymmetric model. 

Very interesting is also the electric dipole moment of the neutron. It has been
studied experimentally at the ILL in Grenoble and at Gatchina in St. Petersburg.
The present value lies at 

\noindent
\begin{equation}
d_n \lesssim 10^{-26} e \cdot cm\,.
\label{eq:3}
\end{equation}

It is easy to see that an electric dipole moment requests violation of time
reversal symmetry and of parity (see fig. 8). 

\begin{figure}[h]
\centerline{\epsfxsize=3.9in\epsfbox{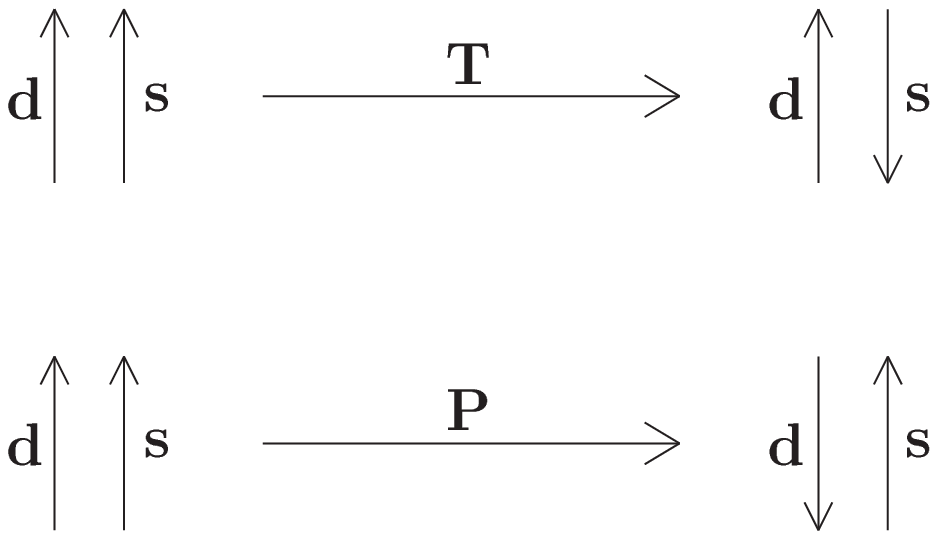}}   
\caption{Electric dipole moment of the neutron. Since the spin of
the neutron is the only specific direction the dipole moment has to be parallel or
antiparallel to the spin. Time reversal does not change the direction of the
dipole moment, but the direction of the spin and thus time reversal symmetry
requests a zero electric dipole moment (see upper part of the figure). The
parity operation does change the direction of the dipole moment, but not the
direction of the spin. The lower part of the figure shows that conservation of
parity requests a zero electric dipole moment of the neutron. The CPT theorem
indicated that time reversal symmetry must be violated if CP is not a good
symmetry. Violation of parity is well-known. Since time reversal and parity are
violated, there must be at some level an electric dipole moment of the neutron.
The present upper limit is given in equation 3 ($10^{-26} e \cdot cm$). 
The electroweak standard model
predicts an electric dipole moment of the neutron of the order of $10^{-32} e
\cdot cm$. But extensions of the standard model predict much larger electric dipole
moments of the neutron. Improving the upper limit for the electric dipole moment
of the neutron means to exclude some of these models which go beyond the
standard model. \label{inter}}
\end{figure}


\section{Examples from different fields}

\subsection{The perturbative chiral quark model (P$\chi$QM)}

The Lagrangian of Quantum Chromodynamics (QCD) is invariant under the 
$SU(3)_L$ and $SU(3)_R$ flavour transformation for the first three quarks: u, d
and s. If the current masses of this quarks are put equal to zero. 

\noindent
\begin{eqnarray}
\mathcal L_{QCD} & = & \sum_f \bar{\Psi}_f \left( i D\!\!\!\!/ - \hat{m}_f \right)
\Psi_f\ -\ \frac{1}{4}\ F_{\mu \nu}^a\ F_a^{\mu \nu}\nonumber \\ 
with\ \ \ D\!\!\!\!/ & = & \gamma^{\mu} \left( \theta_{\mu} + \frac{i}{2}\ g\
\lambda_a\ A^a_{\mu}\right)\\ 
F_{\mu \nu}^a & = & \partial_{\mu}\ A_{\nu}^a\ -\partial_{\nu}\ A_{\mu}^a\ - g\
f_{abc}\ A_{\mu}^b\ A_{\nu}^c \nonumber\,.
\label{eq:4}
\end{eqnarray}

This means that the right-handed quarks (index R) stay always right-handed and left-handed
quarks stay (index L) always left-handed in the chiral limit. 

\begin{eqnarray}
q_{L/R} & = & \frac{1}{2} \left( 1 \mp \gamma^5\right) \left( \begin{array}{c}  u \\ d\\
s \end{array} \right) \nonumber\\
m_f\ \bar{\Psi}_f\ \Psi_f & = & \hat{m}_f \left[\bar{\Psi}_{fL}\ \Psi_{fR}\ +\
\bar{\Psi}_{fR}\ \Psi_{fL} \right] \, .
\label{eq:5}
\end{eqnarray}

The lower index f runs over the three flavours up u, down d and strange s. The
indices a, b and c run over the three colors and $f_{abc}$ is the structure
constant of SU(3), and $\lambda_a$ are the Gell-Mann color matrices. 
The second
part of eq. (5) shows that the mass term with finite current masses of the
quarks $m_f$ violates chiral symmetry and scatters right-handed into left-handed
and left-handed into right-handed quarks. 

The perturbative chiral quark model \cite{pqm,xqm,pxqm} uses like chiral
perturbation theory the non-linear $\sigma$ model to restore chiral symmetry
even under the presence of a confining scalar S(r) and perhaps also vector
potential \cite{bhad,thom}. The chiral perturbation theory \cite{leut} eliminates
quarks and gluons and describes everything  on the level of hadrons and the
pseudoscalar Goldstone bosons. It has been shown that this approach is
equivalent in a low energy limit to QCD as an effective field theory if the free
parameters are adjusted accordingly. The perturbative chiral quark model 
(P$\chi$QM) can be shown  to be equivalent to a description on the hadron level
including all hadrons \cite{wein,weinb,hadji,sal}, if one chooses form factors
of the hadrons for the decomposition into quarks, which are Lorentz and Gauge
invariant and fullfill the Ward identities. A proof, that the P$\chi$QM is at
low energies equivalent to QCD, does not exist.

If one expands the effective Lagrangian of the P$\chi$QM up to second order in the
pion field and restricts the expression to SU(2)-flavour, one obtains: 

\begin{eqnarray}
\mathcal L_{P \chi QM} & = & \mathcal L_0 + L_{int} + 0 \left(\pi^3\right) \nonumber \\
\mathcal L_0 & = & \bar{\Psi}_q\ \left\{ i\ \partial\!\!\! /  - S (r) -
\gamma^{\circ}\ V (r) \right\} \Psi_q \nonumber \\
& & - \frac{1}{2}\ \vec{\pi} \left\{ \partial_{\mu}\
\partial^{\mu} + m^2_{\pi} \right\} \vec{\pi} (x)\\
\mathcal L_{int} & = & - \frac{1}{f_{\pi}}\ S (r)\ \bar{\Psi}_q\ i\ \gamma^5 \left(
\vec{\tau} \cdot \vec{\pi} \right) \Psi_q \nonumber\\
& & + \frac{1}{2f^2_{\pi}}\ S (r)\ \bar{\Psi}_q\ \vec{\pi} \cdot \vec{\pi}\ \Psi_q
\nonumber\\
\Psi_q & = & \sum_{\alpha}\ b_{\alpha}\ u_{\alpha} (x) + \sum_{\beta}\ d^+_{\beta}\
v_{\beta} (x) \nonumber\, .
\label{eq:sp}
\end{eqnarray}

In this expression we replaced the form factors for the decomposition of the
nucleons into quarks by a scalar S(r) and  a vector V(r) confining potential.
This naturally violates Lorentz and Gauge invariance. But the model is
presently improved by removing these potentials and having instead Lorentz and
Gauge invariant form factors, which connect the hadrons with the quarks. The
quark wave function $\psi_q$ is calculated by solving the Dirac equation and
one obtains bound state quark wave functions $u_{\alpha} (x)$ and the
corresponding quark annihilation operator $b_{\alpha}$. At the same time one
obtains also the wave function for the antiquarks $(v_{\beta})$. But due to the
presence of the vector potential V(r), which changes sign from quarks to
antiquarks, the model cannot be applied to calculate antiquark and thus also
antinucleon wave functions. This difficulty does not exist if one introduces
Lorentz and Gauge invariant form factors for the decomposition of the hadrons
into quarks. 

The first term of the interaction Lagrangian  $\mathcal L_{int}$ in eq. (6) is the usual
pseudoscalar coupling  of the pions to quarks. One can show by a unitary
transformation that this term can be transformed into the pseudovector
coupling. The two formulations are completely equivalent \cite{pseudo}.The
second interaction term is the so-called Seagull term, which is obtained by the
expansion up to second order of the non-linear $\sigma$ model. The original
Lagrangian of the P$\chi$QM fulfils the Gell-Mann Oaks Renner and the
Gell-Mann Okubo relations. 

\begin{eqnarray}
m^2_{\pi} & = & 2 \hat{m}\ B\ ;\ m^2_K = (\hat{m} + \hat{m}_s) B\nonumber\\
m^2_{\eta} & = & \frac{2}{3} (\hat{m} + 2 m_s) B\nonumber\\
3 m^2_{\eta} + m^2_{\pi} & = & 4 m^2_K \nonumber\\
with\ :\ B & \equiv & - \langle 0\ |\ \bar{u}\ u\ |\ 0 \rangle = m^2_{\pi} / (2
\hat{m})\\
& = & 1.4\ \left[GeV \right]\nonumber\\
\hat{m} & = & \frac{1}{2}\ (\hat{m}_u + \hat{m}_d) = 7 MeV\nonumber\\
\hat{m}_s & = & 175 MeV \nonumber\, .
\label{eq:7}
\end{eqnarray}

The scalar and vector confining potential S(r) and V(r) are parametrized by the
solution of the Dirac equation in an oscillator potential by two parameters,
which are fitted to the axial vector coupling constant $g_A = 1.25 $ and to the
mean squared radius of the quark content of the nucleon $\langle r^2_{charge}
\rangle = 0.6 \pm 0.1  [fm^2]$. 

\begin{equation}
q_{1s} (\vec{r}) = N\ exp \left[ - \frac{r^2}{2 R^2} \right] \left( \begin{array}{cc} 1 & \\
i \rho & \frac{\vec{\sigma} \cdot \vec{r}}{R} \end{array} \right) \chi_s\ \chi_f\ \chi_G\,.
\label{eq:8}
\end{equation}

The two parameters fitted are R, which parametrizes the radius of the quark
content of the nucleon and $\rho$ which determines the importance of the two
small relativistic components. $\rho$ is adjusted to the axial coupling constant. 

\begin{equation}
g_A = \frac{5}{3}\ \left[1 - \frac{ 2 \rho^2}{1 + \frac{3}{2} \rho^2} \right] = 1.25\,.
\label{eq:murn}
\end{equation}

Here I want to show the application of this model to the pion nucleon
$\Sigma$-term, including SU(3) flavour. 

\begin{eqnarray}
\sigma^{\pi}_N & \equiv & \hat{m}\ \langle p\ |\ \bar{u}\ u\ + \bar{d}\ d\ |\ p
\rangle\nonumber\\
& = & \hat{m}\ \frac{\partial\ m_N}{\partial\ \hat{m}}\nonumber\\
\mathcal H_{\chi SB} & = & \bar{q}\ \mathcal M\ q + \frac{B}{8}\ T_r \left\{ \Phi,\ \left\{
\Phi,\ \mathcal M\right\}_+ \right\}_+\nonumber \\
\mathcal M & = & \left( \begin{array}{ccc} \hat{m} & 0 & 0 \\ 0 & \hat{m} & 0 \\0 & 0 &
m_s \end{array} \right) \nonumber\\
\Phi & = & \left\{ \pi^+;\ \pi^0;\ \pi^-;\ K^+;\ K^0;\ \bar{K^0};\ K^- \right\}\nonumber\\
\hat{m} & = & \frac{1}{2}\ \left( \hat{m}_u + \hat{m}_d \right) = 7\ [MeV]\,.
\label{eq:10}
\end{eqnarray}

The pion nucleon $\Sigma$-term is calculated by introducing the chiral symmetry
breaking ($\chi$SB) Hamiltonian $\mathcal{H}_{\chi SB}$ according to the
diagrams of figure 9 into the quark and the Goldstone boson (dashed lines)
propagator.

\begin{figure}[h]
\centerline{\epsfxsize=3.9in\epsfbox{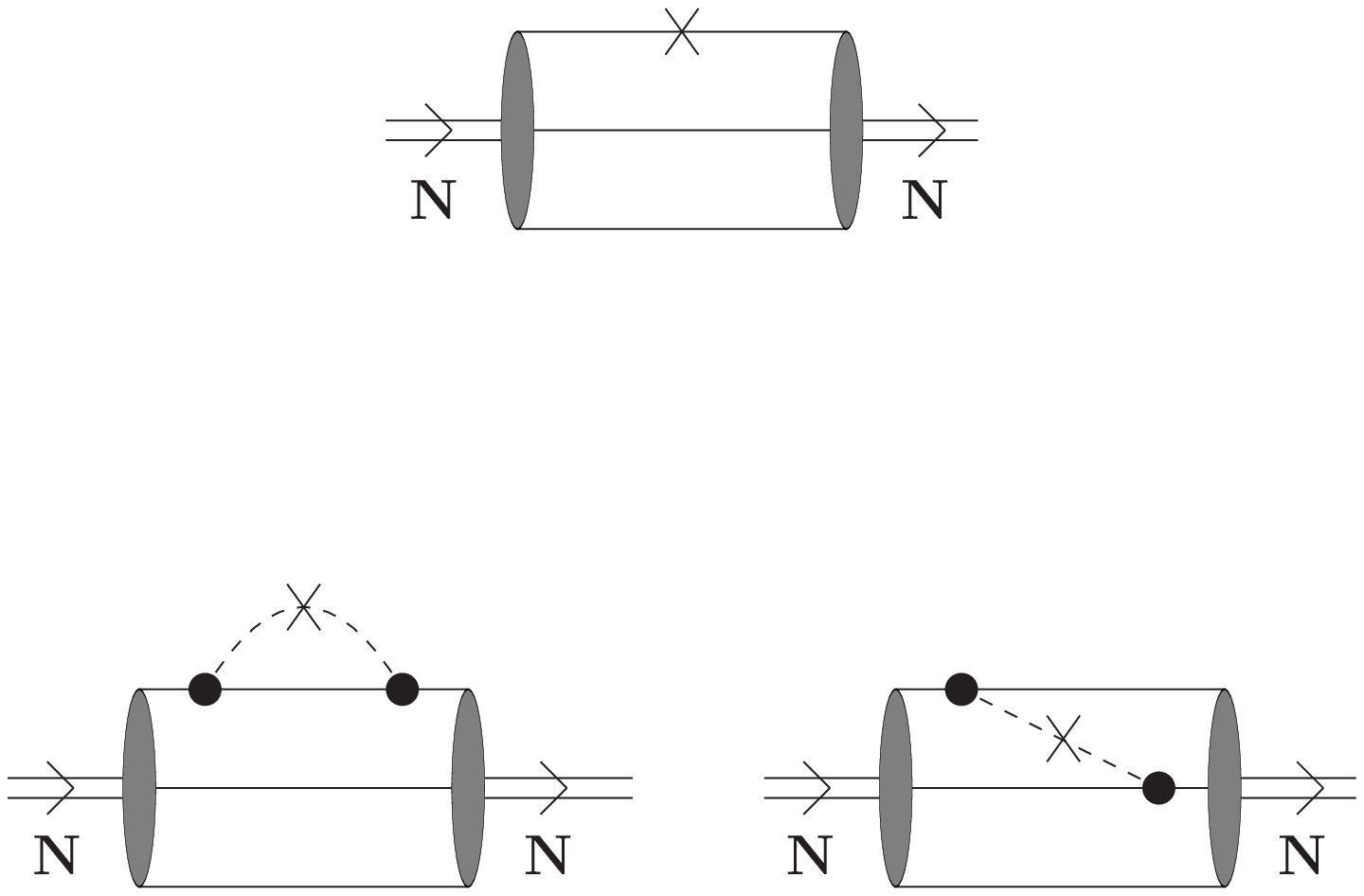}}   
\caption{Diagrams for the  pion nucleon $\Sigma$-term in SU(3)
flavour. The cross indicates the introduction of the chiral symmetry breaking
term $\mathcal{H}_{\chi SB}$ into the quark (solid lines) and the pseudoscalar
Goldstone boson (dashed line: indicating pion, kaon and $\eta$ meson propagators). \label{inter}}
\end{figure}

The results are contained in table 1. The contribution of the valence quarks
is only 13.1 MeV, while the pion cloud contributes the largest part to the
pion-nucleon $\Sigma$ term. The final result is in good agreement with the
extraction of this quantity by J. Gasser from the data (which is controversal). 

\begin{table}[h]
\tbl{ Contribution of the valence quarks $\sigma$(q). The pion-nucleon $\Sigma$ term 
is the sum of the valence contribution $\sigma (q)$, the pion cloud
$\sigma$($\pi$), the kaon cloud $\sigma$(K) and the $\eta$ cloud $\sigma$($\eta$). 
The theoretical error of $\sigma$(total) 
is due to the uncertainty in the mean square radius of the quark
content of the nucleon $\langle r^2 \rangle = 0.6 \pm 0.1 [fm^2]$.\vspace*{1pt}}
{\footnotesize
\tabcolsep35pt
\begin{tabular}{|c|c|}
\hline
{} & $MeV$ \\[1ex]
\hline
$\sigma (q)$ &13.1 \\[1ex]
\hline
$\sigma (\pi)$ &30.2 \\[1ex]
$\sigma (K)$ &1.7\\[1ex]
$\sigma (\eta)$ &0.08 \\[1ex] 
\hline
$\sigma (total)$ &45 $\pm$ 5\\[1ex]
\hline
$\sigma (exp)$ & $\approx $ 45\\
\hline
\end{tabular}\label{tab1} }
\vspace*{-13pt}
\end{table}


Different contributions of the valence quarks q, of the pion cloud $\pi$, of the
kaon cloud K and the eta cloud $\eta$ to the pion-nucleon $\Sigma$ term. The
sum is compared with the experimental value, extracted by J. Gasser from the
data.

\subsection{Nuclear Structure and VAMPIR}

The standard approach to describe the structure of light nuclei is presently the
shell model, while in heavier nuclei one uses mainly the Quasi-Particle Random
Phase Approximation (QRPA). 

Apart of these approaches one uses in light nuclei up to about mass number 10
the Green's function Monte Carlo method \cite{gran}. In heavier nuclei one can
use the Monte Carlo shell model approach of Otsuka and co-workers \cite{ots}
and the Tuebingen VAMPIR (Variation After Mean field Projection In Realistic model
spaces and with realistic forces) \cite{vamp}. The shell model Monte Carlo method and also the
VAMPIR approach select very carefully the configurations which are included in
the diagonalization to find the many-body nuclear wave function. In this way
the VAMPIR approach can obtain with three to five projected
Hartree-Fock-Bogoliubov configurations the same quality as the shell model with
several million configurations. The VAMPIR approach starts from the quasi-particle
transformation and the intrinsic Hartree-Fock-Bogoliubov wave function. 

\begin{eqnarray}
\alpha^+_i & = & \sum_a\ \left\{ A_{ai}\ c^+_a\ + B_{ai}\ c_a \right\}\nonumber\\
|HFB\rangle & \equiv &  | \rangle = \prod_{all\ i}\ \alpha_i\ |\ 0
\rangle\nonumber\\
E^{\pi, Z, N}_{JM} & = & \frac{ \langle\ HFB\ |\ \hat{H}\ \hat{P}_{JM}\
\hat{P}_{\pi}\ \hat{P}_z\ \hat{P}_N\ |\ HFB \rangle}{\langle\ HFB\ |\ \
\hat{P}_{JM}\ \hat{P}_{\pi}\ \hat{P}_Z\ \hat{P}_N\ |\ HFB\rangle}\nonumber\\
& = & f \left( A_{ai},\ B_{ai} \right)\ =\ Minimum\, .
\label{eq:11}
\end{eqnarray}

The first Hartree-Fock-Bogoliubov (HFB) configuration is found by minimizing the
energy $E_{JM}$ after projection onto good symmetries like angular momentum,
angular momentum projection, parity, time reversal symmetry, proton number and
neutron number as a function of the coefficients in the quasi-particle
transformation $A_{ai} $ and $ B_{ai}$. As more symmetries one breaks in the 
quasi-particle transformation from the particle creation and annihilation operators
$c_a^{\dagger}$ and $c_a$ as better the final wave function will be. But one
has to project the HFB wave function on all good symmetries before minimization
of the energy. Further configurations are found by requesting orthogonality of
the new configurations with the previous ones by Schmidt orthogonalization. 

\begin{figure}[h]
\centerline{\epsfxsize=4.3in\epsfbox{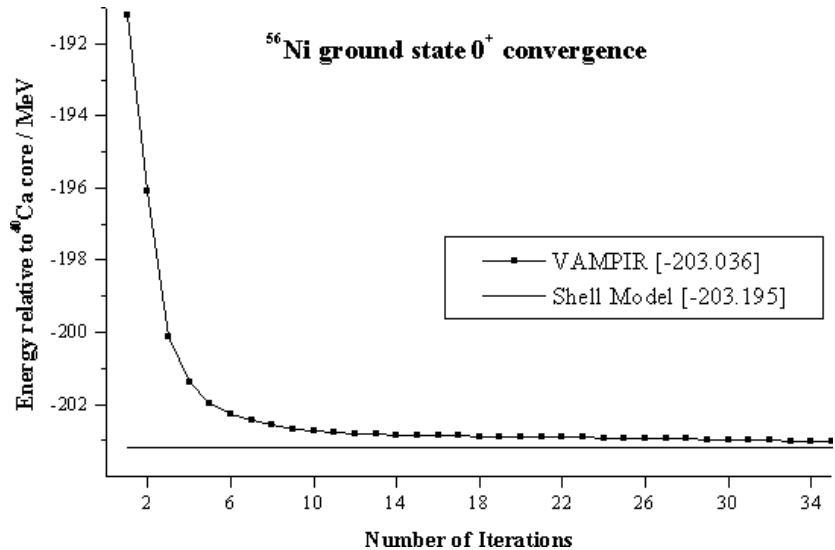}}   
\caption{Bound state $0^+$ binding energy of $^{56}Ni$ relative to $^{40}Ca$ 
convergence of the VAMPIR
approach with a single configuration as in eq. (11) compared with a shell model
diagonalization (solid straight line) with more than 15 million configurations
of the Strassburg-Madrid group. \label{inter}}
\end{figure}

Figure 10 shows the convergence of the minimization of the energy of the $0^+$
state (11) of $^{56}Ni$ compared with the shell model calculation of more than
15 million configurations. In the VAMPIR approach one has only one
configuration, but the model space is chosen to be the pf-shell as in the shell
model approach. In both calculations the same Hamiltonian is used. 

An usual restriction in the shell model calculations for the pf-shell nuclei is
to allow in the $f_{7/2}$ shell for the protons and the neutrons only six holes
(maximum number of holes would be 8). If one uses in the VAMPIR approach two
configurations, one obtains a lower energy than the shell model approach with
this truncation. With five configurations in a VAMPIR approach we reach a lower
energy than the shell model Monte Carlo approach \cite{ots} with 30
configurations. 

The VAMPIR approach is able to use a larger basis than the pf-shell and in this
case one reaches a lower energy than the best shell model calculations with up to 
100 million configurations, which must be restricted to the
pf-shell. 

Thus one sees it pays to select carefully the configurations which one is
including into the nuclear structure calculations. The shell model needs so
many configurations, because one takes all possible configurations in a
non-optimal single particle and many-body basis. 

\subsection{ Nuclear Astrophysics at the Gamow Peak }

Nuclear reactions relevant for the formation of the elements in the stars
are far below the Coulomb barrier. A very intersting
example is the fusion reaction

\noindent
\begin{equation}
^3He + ^3He\ \rightarrow\ ^4He + 2p\,,
\label{eq:murn}
\end{equation}

which essentially determines the high energy neutrino flux produced by the sun. Since the
reaction is extremely weak at the temperatures of about 15 million degrees in
the sun, one measures the reaction cross-section in the laboratory at higher
energies and extrapolates with the astrophysical S-factor down to the energy
(Gamow peak), where the reaction really happens in the sun. 

A collaboration from Napoli and Bochum was now able to measure for the first
time in the Gran Sasso this reaction at the Gamow peak (see figure 11). 

\begin{figure}[h]
\centerline{\epsfxsize=3.5in\epsfbox{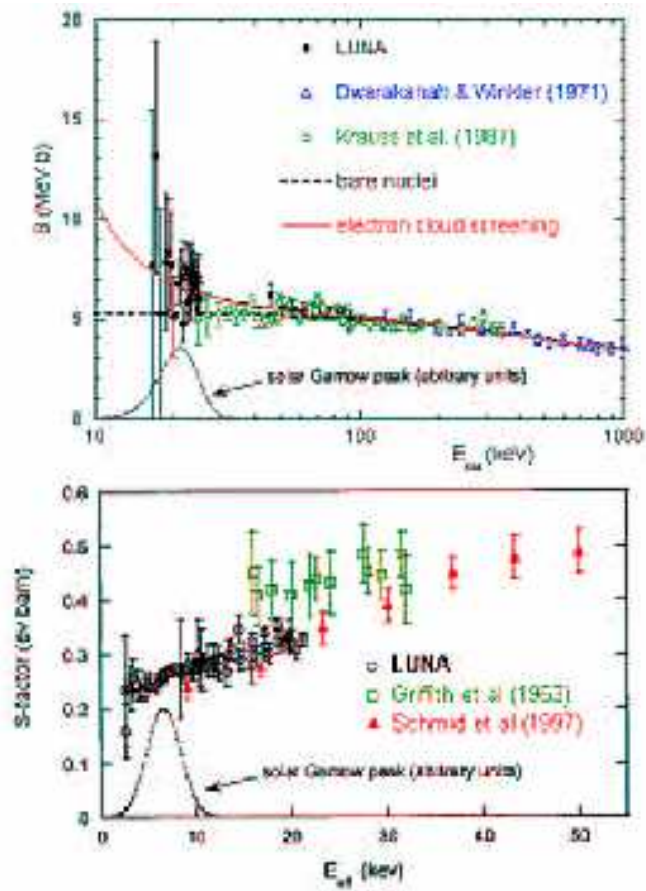}}   
\caption{Cross-section of the fusion reaction $^3He + ^3He$ of
eq. (12) measured by the Luna collaboration in the Gran Sasso over 12 order of
magnitudes until down to the Gamow peak. \label{inter}}
\end{figure}

\subsection{The Solar Neutrino Problem}

In the sun hydrogen is burning into helium

\noindent
\begin{eqnarray}
2 e^- + 4 \cdot p\ \rightarrow\ ^4_2He_2 + 2\nu_e + Energy\nonumber\\
pp-neutrinos\ \ E_{\nu} \le 420 keV\nonumber\\
e^- +^7Be-neutrinos \ \ 380 + 860 keV\nonumber\\
^8B-neutrinos\ \ E_{\nu} \le 14.6 MeV\,,
\label{eq:13}
\end{eqnarray}

in a reaction network, which leads to the production of $^4He$. Neutrinos are
produced  at different places. The main neutrino sources as indicated in eq.
(13) is the proton-proton chain, the electron capture in $^7Be$ and the beta
decay of $^8B$.

Different detectors investigated the neutrino flux from the sun and found all
fewer neutrinos than expected from the standard solar model. The Cl detector in
the Homestake mine of Davies measures mainly the B neutrinos and partially also the
$^7Be$ neutrinos from the 860 keV line. The detector finds about 34 $\%$ of the
neutrinos expected. The two Ga detectors GALLEX in the Gran Sasso and the SAGE
in the Caucasus have a very low threshold of about 230 keV and are mainly
sensitive of the neutrinos from the proton-proton chain and to the two discrete
lines from $^7Be$. These detectors find 59 and 58 $\%$ of the neutrinos
expected. 

The Cherenkov detectors Kamiokande and Super-Kamiokande in Japan, which are
only sensitive to the high energy $^8B$ neutrinos measure 55 and 46 $\%$ of the
neutrinos expected. The breakthrough came from the Cherenkov detector with
heavy water $D_2 O $ in the Sudburry Neutrino Observatory (SNO). SNO measures at
the same time charge and neutral current reactions. It also can separate the neutrino
reactions on nucleons from the ones on electrons. By using for the charge
current reactions on electrons the precise measurement by Super-Kamiokande one
can separately determine the total neutrino and the electron
neutrino flux from the sun. This separation is only possible by measuring the charge current
reaction on the neutron (see figure 12), which requires heavy water. 

\begin{figure}[h]
\centerline{\epsfxsize=2.8in\epsfbox{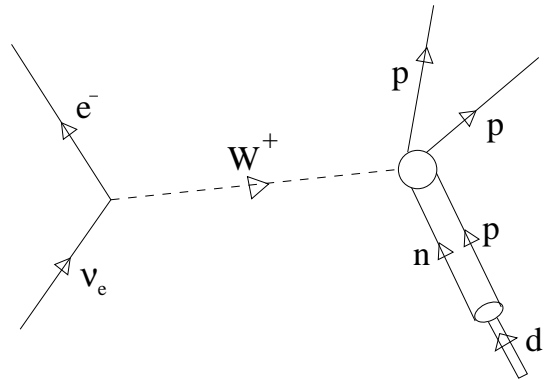}}   
\caption{Charge current reaction of an electron neutrino on the
deuteron. This charge current reaction cannot be measured at the neutrons in
$^{16}O$, since they are too strongly bound. \label{inter}}
\end{figure}

By separating with the angular distribution the charge current reaction on a
neutron from the charge and the neutral current reaction on an electron and by
using the precise measurement of Super-Kamiokande for the charge current
reaction and the neutral current reaction on an electron, one is able to
determine separately the electron neutrino flux and the total neutrino flux from
the sun. The total neutrino flux from the sun is exactly the one expected from
the standard solar model for electron neutrinos. So one concludes that no
electron neutrinos from the sun are missing. They only oscillated in muon or
tauon neutrinos. The SNO data also exclude within the precision of the
measurement (which is not too good) oscillations into sterile neutrinos. 

So finally we have no neutrino puzzle. We must only understand the neutrino
oscillations more in detail. This means we must get to know the neutrino masses
and the neutrino mixing matrix. SNO favours the large mixing angle solution
for solar neutrinos and excludes practically the small mixing angle solution.
So the mixing of the first and the second mass eigenstates is about
$\theta_{12} \cong 35^\circ (\pm 5^{\circ})$ and from the atmospheric neutrinos we find
$\theta_{23} = 45^{\circ}$, while the data seem to suggest $\theta_{13} \cong
0^{\circ}$ (with a large error). $\theta_{13} = 0$ would not allow CP violation in the neutrino
sector. Thus a precise determination of $\theta_{13}$ is very important.

\section{ Conclusion}

It is difficult to make predictions, especially if it
concerns the future. I tried to extrapolate topics of which I think they are
interesting presently and will be interesting also in the next future. 

\begin{enumerate}
\item We shall use more and more lattice QCD and also analytical methods to
solve QCD to determine properties of hadrons. On the other side one must show
that the present effective field theories can be justified by QCD as a low
energy limit like chiral perturbation theory. As a candidate for such an
effective field theory which includes pseudoscalar Goldstone bosons and quarks
I discussed the perturbative chiral quark model (P$\chi$QM), which we developed
in Tuebingen. 

\item Due to the advent of radioactive beams nuclear structure has a
renaissance. As an approach to get better and better solution of the nuclear
structure problem also in heavier nuclei, I discussed VAMPIR \cite{vamp}, which
with very few configurations can obtain the same quality of results as shell
model calculations with several million configurations. 

\item Ultrarelativistic heavy ion collisions will study (and detect) 
in the future more in
detail the phase transition from nuclear to quark matter. 

\item With the radioactive beams we will be able to study more and more
reactions which are relevant for nuclear astrophysics, especially for the
formation of the elements in the stars. Measurements of astrophysical relevant
reactions until down to the Gamow peak are now possible by small accelerators in
underground laboratories. 

\item Neutrino physics is at the moment a very fast developing field. It seems
that the solar neutrino problem is practically solved. One still has to verify
if the data from LSND (Los Alamos Scintilator Neutrino Detector) are
correct. For this the experiment Mini-BooNE is starting to take data at Fermi Lab. We wait for
more data from SNO (Sudburry Neutrino Observatory) and more data from Long
Baseline neutrino oscillation experiments to determine the neutrino mixing matrix. 

\item Especially interesting are tests of new physics beyond the standard model
of Grand Unified  and Supersymmetric theories. Such models can be tested in
rare decays. Among them rates very high the neutrinoless double beta decay, which can distinguish
between Dirac and Majorana neutrinos, and
the measurement of the electric dipole moment of the neutron. 

\end{enumerate}



\begin{thebibliography}{0}
\bibitem{leut} H. Leutwhyler,  {\it Ann. of Physics} (1994).

\bibitem{wein} S. Weinberg, {\it Phys. Rev.} {\bf 130}, 
776 (1963).

\bibitem{weinb} S. Weinberg,
{\it Physica } {\bf 96 A}, 327 (1979).

\bibitem{hadji} D. Hadjimichel, G. Krein, S. Szpigel, J. S. Da Veiga, 
{\it Ann. of Physics} {\bf 268}, 105 (1998).

\bibitem{sal} A. Salam, {\it Nuovo Cimento} {\bf 25}, 224 (1962).

\bibitem{gu} M. Strohmeier, Th. Gutsche, A. Faaessler, R. Vinh Mau,  {\it Phys. Lett. }{\bf B438},
21 - 26 (1998);

\noindent
{\it Phys. Rev.}{\bf D 60}; 010 (1999);

\noindent
{\it Acta Phys. Polon.}{\bf B31} 2657 (2000);

\noindent
{\it Nucl. Phys.} {\bf A684} 345 (2001).

\bibitem{am} C. Amsler, {\it Phys. Lett. }{\bf B 541}, 22 (2002).

\bibitem{NN-Quark} A. Faessler, F. Fernandez, {\bf Phys. Lett. }{\bf 124 B}, 145 (1983);

\noindent
K. Br\"auer, A. Faessler, F. Fernandez, K. Shimizu, {\it Z. Phys. }{\bf A 320}, 609 (1985);

\noindent
A. Valcarse, A. Buchmann, F. Fernandez, Y. Yamauchi, A. Faessler, {\it Phys. Rev.} {\bf C 50}, 2246
(1994);

\noindent
A. Valcarse, A. Faessler, F. Fernandez, {\it Phys. Lett. }{\bf B 345}, 367 (1995) and {\it Phys. Rev.
}{\bf C 51}, 1480 (1995).

\bibitem{kukulin} A. Faessler, V. I. Kukulin, I. t. Obukhovsky, V. N. Pomerantsev, {\it J.
Phys.}{\bf G 27}, 1851 (2001).

\bibitem{doenau} S. Frauendorf, {\it Rev. Mod. Phys.}{\bf 73}, 463 (2001).

\bibitem{huebel} R. M. Clark et al., {\bf A 562}, 121 (1993) and S. Frauendorf and F. Doenau, {\it
Proceedings of Oak Ridge Conference on High Angular Momentum, 1982}.

\bibitem{jen} D. J. Jenkins et al., nucl-ex/0007004.

\bibitem{strass} A. Poves, J. Sanches-Solano, E. Caurier, F. Nowacki, {\it Nucl. Phys. }{\bf A 694},
157 (2001) and {\bf A 693}, 374 (2001).

\bibitem{gran} R. B. Wringa, S. C. Pieper, J. Carlson, V. R. Pandharipande, {\it Phys. Rev.} {\bf C
62}, 014001 (2000).

\bibitem{vamp} T. Hjelt, K. W. Schmid, A. Faessler, {\it Nucl. Phys.} {\bf A 697}, 164 (2002);

\noindent
A. Petrovici, K. W. Schmid, A. Faessler, {\it Nucl. Phys. }{\bf A 689},  707 (2001).

\bibitem{ots} M. Honma, T. Otsuka, B. A. Brown, T. Mizusaki, {\it Phys. Rev.} {\bf C 65}, 061301
(2002).

\bibitem{ring} T. Niksik, D. Vretenar, P. Pirelli, P. Ring, {\it Phys. Rev.} {\bf C 66}, 024306
(2002).

\bibitem{berna} R. Bernabei, {\it Prog. Part. Nucl. Phys. }{\bf 48}, 263. (2002).

\bibitem{dark} M. Milgrom, {\it Scientific American}, August 2002.

\bibitem{klap} H. V. Klapdor-Kleingrothaus, A. Dietz, I. V. Krivosheina, {\bf Part. Nucl. Lett.}
{\bf 110}, 57 (2002).

\bibitem{pqm} V. E. Lyubovitskij, T. Gutsche, A. Faessler, R. Vinh Mau, {\it Phys. Lett.} {\bf
B520}, 204 (2001) and {\it Phys. Rev.} {\bf 63}, 054026 (2001).

\bibitem{xqm} V. E. Lyubovitskij et al., {\it Phys. Rev.}{\bf C65}, 025202 (2002). 

\bibitem{pxqm} V. E. Lyubovitskij, P. Wang, T. Gutsche, A. Faessler, hep-ph/0207225 and
hep-ph/0205251.

\bibitem{bhad} R. K. Bhaduri, Models of the Nucleon; Addison Wesley Publishing Company, Redwood
City, 1988.

\bibitem{thom} A. Thomas, W. Weise, The Structure of the Nucleon; Wiley-VCH, Berlin, 2000.

\bibitem{pseudo} V. E. Lyubovitskij, Th. Gutsche, A. Faessler, R. Vinh Mau, {\it Phys. Lett.} {\bf B
520}, 204 (2001) and V. E. Lyubovitskij et al., {\it Phys. Rev.}{\bf C65}, 025202 (2002). 


\bibitem{iva} M. A. Ivanov, V. E. Lyubovitskij, {\it Phys. Rev.}{\bf D 56}, 348 (1997).

\bibitem{ivanov} M. A. Ivanov, M. P. Locher, V. E. Lyubovitskij, {\it Few Body Systems}{\bf 21}, 131
(1996).

\bibitem{sm} A. Poves, J.Sanchez-Solano, E. Caurier, F. Nowacki, nucl-th/0210069.

\bibitem{sno} SNO-Collaboration, {\it Phys. Rev. Lett}{\bf 89}, 011302 (2002) and {\it Phys. Rev.
Lett.}{\bf 89}, 11301 (2002).

\end{thebibliography}
\end{document}